\documentclass{aa}  

\usepackage{graphicx}
\usepackage{txfonts}
\usepackage{subcaption}         
\usepackage{booktabs}

\usepackage{natbib}
\bibpunct{(}{)}{;}{a}{}{,}
\usepackage{natbib,twoopt}
\usepackage[breaklinks=true]{hyperref} 
\usepackage{color}
\definecolor{cobalt}{rgb}{0.06, 0.2, 0.65}
\hypersetup{
  colorlinks,
  citecolor=cobalt,
  linkcolor=[rgb]{0.8, 0.2, 1.0},
  urlcolor=cobalt,
}

\usepackage{multirow} 

\begin{document}

   \title{Probing the environments of FRI and FRII radio galaxies in LoTSS DR2 with galaxy clusters}

 \author{Tong Pan\inst{1}\corrauth{tpan@strw.leidenuniv.nl}
         \and
         Yuming~Fu\inst{1,2}\corrauth{yfu@strw.leidenuniv.nl}
         \and
         H.~J.~A.~Rottgering\inst{1}
          \and
          J.~M.~G.~H.~J. de Jong\inst{1,3}
          \and
          M.~J.~Hardcastle\inst{4}
          \and
          B.~Mingo\inst{4}
          \and
          L.~Clews\inst{5}
          \and
          M.~Magliocchetti \inst{6}
          \and
          J.~W.~Petley \inst{1}
          \and
          Bohan~Yue \inst{1,7}         
         }

  \institute{Leiden Observatory, Leiden University, Einsteinweg 55, 2333 CC Leiden, The Netherlands
             \and Kapteyn Astronomical Institute, University of Groningen, P.O. Box 800, 9700 AV Groningen, The Netherlands
             \and ASTRON, The Netherlands Institute for Radio Astronomy, Postbus 2, 7990 AA Dwingeloo, The Netherlands
             \and Centre for Astrophysics Research, Department of Physics, Astronomy and Mathematics, University of Hertfordshire, College Lane, Hatfield AL10 9AB, UK.      
             \and School of Physical Sciences, Open University, Walton Hall, MK7 6AA, UK.
             \and INAF-IAPS, Via Fosso del Cavaliere 100, 00133, Rome, Italy.
             \and Institute for Astronomy, University of Edinburgh, Edinburgh EH9 3HJ, UK. 
            }

  \abstract
   {}
   {The origin of the morphological dichotomy between Fanaroff-Riley Class I (FRI) and Class II (FRII) radio galaxies has long been debated. In this study, we investigate whether the cluster-scale environment plays a significant role in shaping the morphology of FRIs and FRIIs.} 
   {Using the new morphologically classified catalogue from the second data release of the LOFAR Two-metre Sky Survey (LoTSS DR2), we construct two samples at $z<0.4$: a volume-limited sample with \(L_{144} > 4 \times 10^{24}\) $\mathrm{W\,Hz^{-1}}$ and a paired sample in which each FRII is matched to the nearest FRI in luminosity and redshift. We cross-match these samples with a recent galaxy cluster catalogue based on the DESI Legacy Imaging Survey. A galaxy is considered associated with a cluster with $M_{500} > 0.47 \times 10^{14}M_{\odot}$ if the redshift difference between the galaxy and the cluster centre is below 0.01, and the projected distance between the two is smaller than 2$R_{500}$. }
   {In the volume-limited sample, \(48.6^{+1.8}_{-1.8}\%\) of FRIs and \(30.6^{+2.5}_{-2.3}\%\) of FRIIs are associated with clusters. In the paired sample, \(45.6^{+3.1}_{-3.1}\%\) of FRIs and \(32.6^{+3.0}_{-2.8}\%\) of FRIIs are associated with clusters. This difference in cluster match fractions between FRIs and FRIIs in both samples becomes more pronounced at higher radio luminosities (\(L_{144\mathrm{MHz}} > 10^{26}\,\mathrm{W\,Hz}^{-1}\)). In the volume-limited sample, \(55.6^{+9.2}_{-9.6}\%\) of luminous FRIs and \(19.0^{+5.7}_{-4.6}\%\) of luminous FRIIs are associated with clusters. In the paired sample, \(50.0^{+12.9}_{-12.9}\%\) of luminous FRIs and \(6.7^{+9.5}_{-3.9}\%\) of luminous FRIIs are associated with clusters. 
   Nevertheless, those FRIs and FRIIs that are associated with clusters show similar properties. In particular, the distributions of radio luminosity and stellar mass as functions of cluster richness and \(M_{500}\) are similar for FRIs and FRIIs. The radial density profiles of cluster-associated FRIs and FRIIs both peak at \(0.5 \times R_{500}\) and decline beyond \(R_{500}\), showing very similar spatial distributions within clusters. 
   Furthermore, in the volume-limited sample, \(74.8^{+2.2}_{-2.3}\%\) of cluster-associated FRIs and \(61.9^{+4.4}_{-4.6}\%\) of cluster-associated FRIIs are identified as brightest cluster galaxies (BCGs). In the paired sample, \(78.1^{+3.5}_{-4.0}\%\) of cluster-associated FRIs and \(65.9^{+4.9}_{-5.3}\%\) of cluster-associated FRIIs are identified as BCGs. The median properties of these FRI-BCGs and FRII-BCGs, as well as those of their host clusters, do not show significant differences.
}
   {Compared to FRIs, FRIIs are less frequently found in galaxy clusters, particularly at high radio luminosities. This pattern may be explained by the jet disruption scenario, in which the dense gas in galaxy clusters can slow down or disturb radio jets, making it more difficult for powerful FRII structures to form or remain stable. However, when FRIs and FRIIs do reside in clusters, they appear to inhabit similar environments in terms of richness, mass, and radial position. These findings suggest that, while the cluster-scale environment may influence the cluster match fraction, the morphological distinction between FRIs and FRIIs is unlikely to be primarily driven by cluster-scale properties alone. Improved estimates of cluster richness and mass will help to further clarify the extent to which the cluster environment influences the radio morphology of FRIs and FRIIs.}


    \maketitle
    \nolinenumbers

\section{Introduction}

    Fanaroff-Riley Class I (FRI) and Class II (FRII) radio galaxies, first classified by \citet{Fanaroff1974}, represent two distinct morphological categories of radio galaxies. The FRIs exhibit core-brightened jets that dissipate into diffuse plumes, while the FRIIs display edge-brightened lobes terminating at the hotspots. \citet{Fanaroff1974} also identified a luminosity break between the two classes: FRIIs tend to have radio luminosities above $\sim10^{26}$ W Hz$^{-1}$ at 144 MHz, whereas FRIs generally lie below this threshold.

    For decades, the origin of the structural dichotomy between FRIs and FRIIs has been a subject of active investigation. The morphological differences cannot be attributed solely to the properties of the central engine, as this fails to explain why jets with the same power can develop different morphologies \citep{Hardcastle2020NewAR..8801539H, Mingo2019}. Some models suggest the difference is due to the interplay between the jet and its environment. The edge-brightened FRII radio galaxies are thought to have jets that remain relativistic throughout, terminating in an internal shock visible as a hotspot, while the centre-brightened FRIs are known to have initially relativistic jets that decelerate on kiloparsec scales \citep{Bicknell1995,Laing2002,Tchekhovskoy2016MNRAS.461L..46T,laing2014MNRAS.437.3405L,Kaiser2007MNRAS.381.1548K,perucho2007MNRAS.382..526P}. This jet disruption model provides a physical explanation of how jets with the same power can develop different structures: in low-density environments, jets remain relativistic and well collimated, whereas in denser environments they decelerate, entrain the interstellar medium (ISM), and expand into turbulent FRI plumes. \citet{Ledlow1996AJ....112....9L} found that the FRI and FRII break radio luminosity increases with the increasing optical luminosity of the host galaxy. They found that many FRIs are seen at radio powers equivalent to those of FRIIs, but in host galaxies that are one to two magnitudes brighter. This observation suggests that the density of the host galaxy's ISM is a key factor in shaping radio jet morphology, as brighter galaxies are assumed to have a denser ISM. However, this result was based on strongly flux-limited samples, in which FRIs and FRIIs differed substantially in redshift and environment, introducing serious selection biases. As a result, several studies have questioned whether this relation holds across the full radio galaxy population \citep[e.g.][]{Best2009AN....330..184B, Lin2010ApJ...723.1119L, Wing2011AJ....141...88W, Singal2014MNRAS.442.1656S, Capetti2017A&A...601A..81C, Shabala2018MNRAS.478.5074S,clews2025MNRAS.tmp..936C}.

    Further studies have extended the investigation on FRIs and FRIIs to cluster-scale environments. Some works have found that FRIs tend to reside in richer galaxy clusters than FRIIs at low to intermediate redshifts \citep[e.g.][]{Zirbel1997,Gendre2013MNRAS.430.3086G,Chiaberge2009ApJ...696.1103C,Castignani2014ApJ...792..113C,Magliocchetti2022A&ARv..30....6M}. Some recent works, however, have suggested that the cluster-scale environments of FRIs and FRIIs may not differ as significantly as was previously thought \citep[e.g.][]{Massaro2019ApJS..240...20M,Massaro2020ApJS..247...71M}. One possible reason for the inconsistent conclusions is the selection bias of previous samples. Many earlier studies relied on small, bright samples from surveys such as the Revised Third Cambridge Catalogue \citep[3CR;][]{Smith1976PASP...88..621S} and the VLA Faint Images of the Radio Sky at Twenty centimeters (FIRST) Survey \citep{Becker1994ASPC...61..165B}, which are biased towards luminous sources and under-represent low-luminosity FRIIs. These objects, known as `FRII-Lows', are radio galaxies presenting an FRII morphology but with radio luminosities below $L_{144} < 10^{26}$ W Hz$^{-1}$ \citep{Mingo2019, Mingo2022MNRAS.511.3250M}. FRII-Lows were largely overlooked in early studies due to selection effects, but are now recognised as critical for understanding the full evolutionary path of radio galaxies. In addition, many environmental studies lacked homogeneous morphological classification or robust environmental diagnostics, especially on megaparsec scales.

    To revisit whether the cluster-scale environment influences the morphology of FR radio galaxies, we combine two newly available datasets in this work: the FRI and FRII sample from the LOFAR Two-Metre Sky Survey Data Release Two \citep[LoTSS DR2;][]{Shimwell2022}, as classified by \citet{clews2025MNRAS.tmp..936C}, and the galaxy cluster catalogue from \citet{Wen2024}. The sample from \citet{clews2025MNRAS.tmp..936C} includes 2354 FRIs and 3590 FRIIs identified via visual inspection and machine learning. The \citet{Wen2024} catalogue, based on DESI Legacy Surveys DR9 and DR10, provides positions, redshifts, richness, and mass estimates for 1.58 million galaxy clusters. By cross-matching the two datasets, we enable an accurate comparison of the environments surrounding FRIs and FRIIs. Specifically, we measure cluster match fractions, radial offsets, and richness distributions, while controlling for redshift and radio luminosity.

    Importantly, our sample includes a large number of FRII-Lows, enabling the first statistical study of their cluster-scale environments. This allows us to investigate whether low-luminosity FRIIs differ from their high-luminosity counterparts, whether FRIIs preferentially reside in rich clusters, and to what extent the FR morphological divide is shaped by environment versus intrinsic active galactic nucleus (AGN) properties.

    This paper is structured as follows. Section \ref{sec2} describes the selection of FRI and FRII samples and the cross-matching procedure used to identify their associations with galaxy clusters. Section \ref{sec3} presents the results on cluster match fractions and compares the environmental properties of FRIs and FRIIs. Section \ref{sec4} discusses the implications of the observed environmental similarities for understanding the FRI and FRII dichotomy. Section \ref{sec5} summarises our main conclusions. Throughout this paper, we adopt a flat $\Lambda$ cold dark matter cosmology with $H_0 = 70~\mathrm{km~s^{-1}~Mpc^{-1}}$ and $\Omega_{\mathrm{M}}=0.3$.

\section{Samples and method} \label{sec2}
\subsection{FRI and FRII samples}

    The data used in this work are based on the LOFAR Two-Metre Sky Survey (LoTSS), a high-resolution, low-frequency radio survey conducted with the LOFAR telescope \citep{Shimwell2017}. Its second data release \citep[LoTSS-DR2;][]{Shimwell2022} covers 27\% of the northern sky, providing images with a median root-mean-square (rms) noise of 83~$\mu$Jy/beam at a resolution of 6$''$. Approximately 15\% of the detected sources have been identified as AGNs \citep{Hardcastle2025MNRAS.539.1856H}. Building on this AGN catalogue, \citet{clews2025MNRAS.tmp..936C} applied a newly developed automated classification method based on ridgelines \citep{Barkus2022MNRAS.509....1B}, further refined through visual inspection. This approach resulted in a high-quality sample of radio galaxies, including 2354 FRI and 3590 FRIIs.

\begin{figure}[h]
    \centering

    \includegraphics[width=1\linewidth]{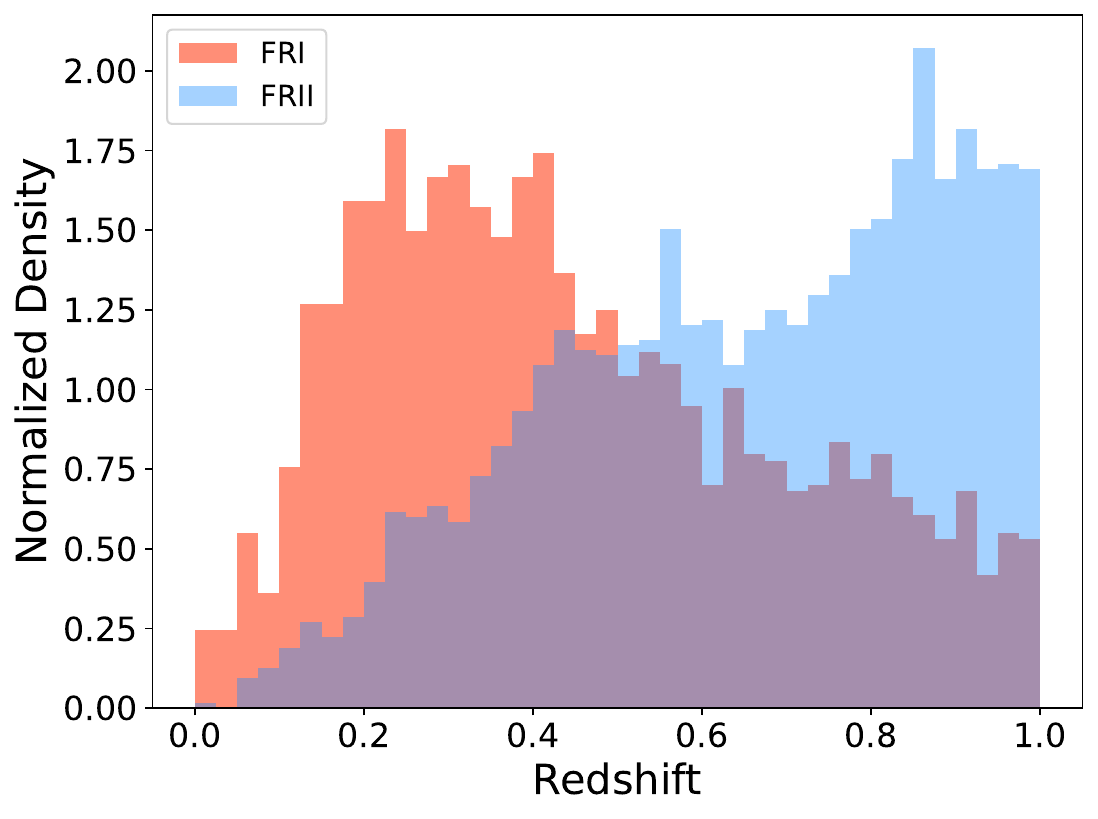}
    \caption{Normalised redshift distribution of FRIs and FRIIs in the redshift range $0 < z < 1$. Each histogram is normalised to the unit area by dividing the bin counts by the total number of sources and the bin width.}
    \label{fig:z_dis}

\end{figure}

\begin{figure}[h]
    \centering

    \includegraphics[width=1\linewidth]{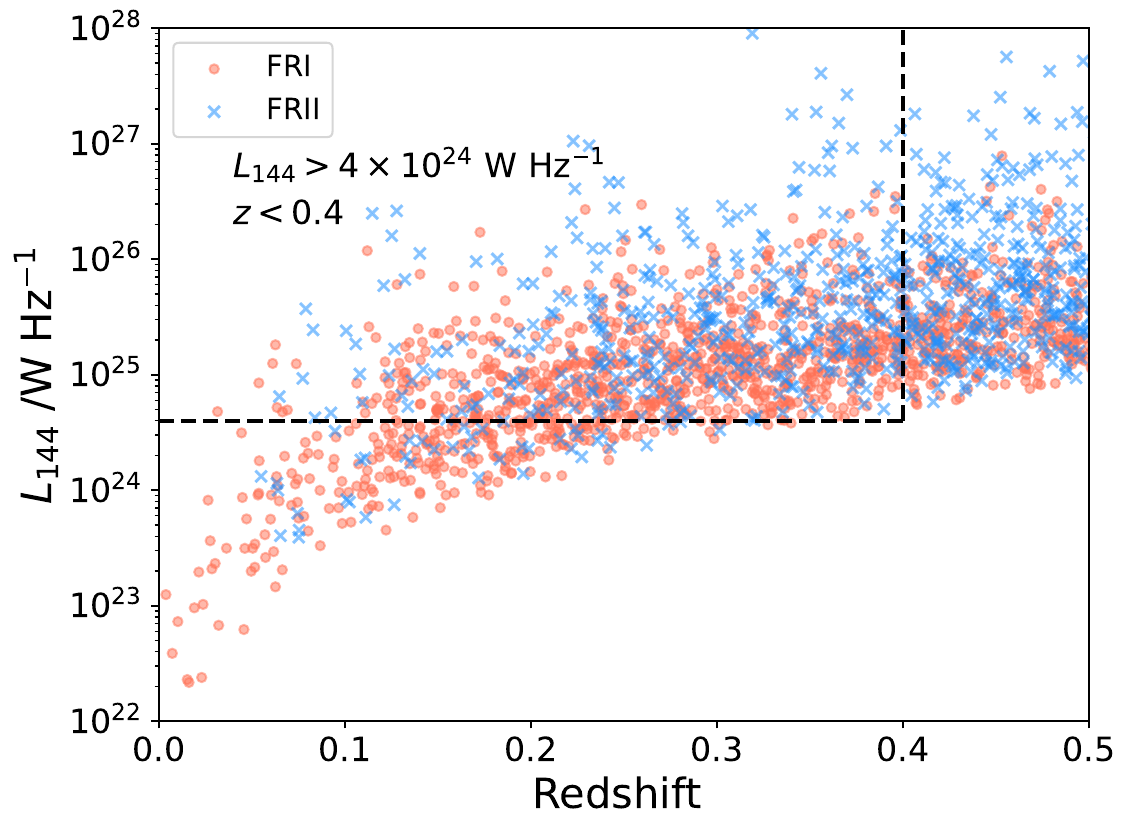}
    \caption{Distributions of radio luminosity at 144 MHz and redshift (\(z\)) for FRIs (red circles) and FRIIs (blue crosses). The vertical dashed line at \(z < 0.4\)) and the horizontal dashed line at \(L_{144} > 4 \times 10^{24}\) $\mathrm{W\,Hz^{-1}}$ define a complete, volume-limited sample of radio sources.}
    \label{fig:pz}

\end{figure}

    Figure~\ref{fig:z_dis} presents the redshift distributions of the FRI and FRII samples within the redshift range $z < 1$. The two populations show broadly different redshift trends: FRIs are most commonly found at lower redshifts than FRIIs, most likely due to the difficulty of detecting fainter sources at high redshift \citep{clews2025MNRAS.tmp..936C}. The number of FRIs declines beyond $z \approx 0.4$, while FRIIs become increasingly dominant at higher redshifts. To mitigate the increasing selection bias caused by flux limits at higher redshifts, we defined a volume-limited sample by selecting sources with $z < 0.4$ and 144 MHz radio luminosity ($L_{144}$ $> 4 \times 10^{24} \,\mathrm{W\,Hz^{-1}}$) based on the radio luminosity-redshift ($L$-$z$) diagram (see Fig. \ref{fig:pz}). Within this selected range, the sample consists of 767 FRIs and 369 FRIIs. Among them, 532 FRIs have spectroscopic and 162 FRIIs have spectroscopic redshifts.

\begin{figure}[h]
    \centering

    \includegraphics[width=1\linewidth]{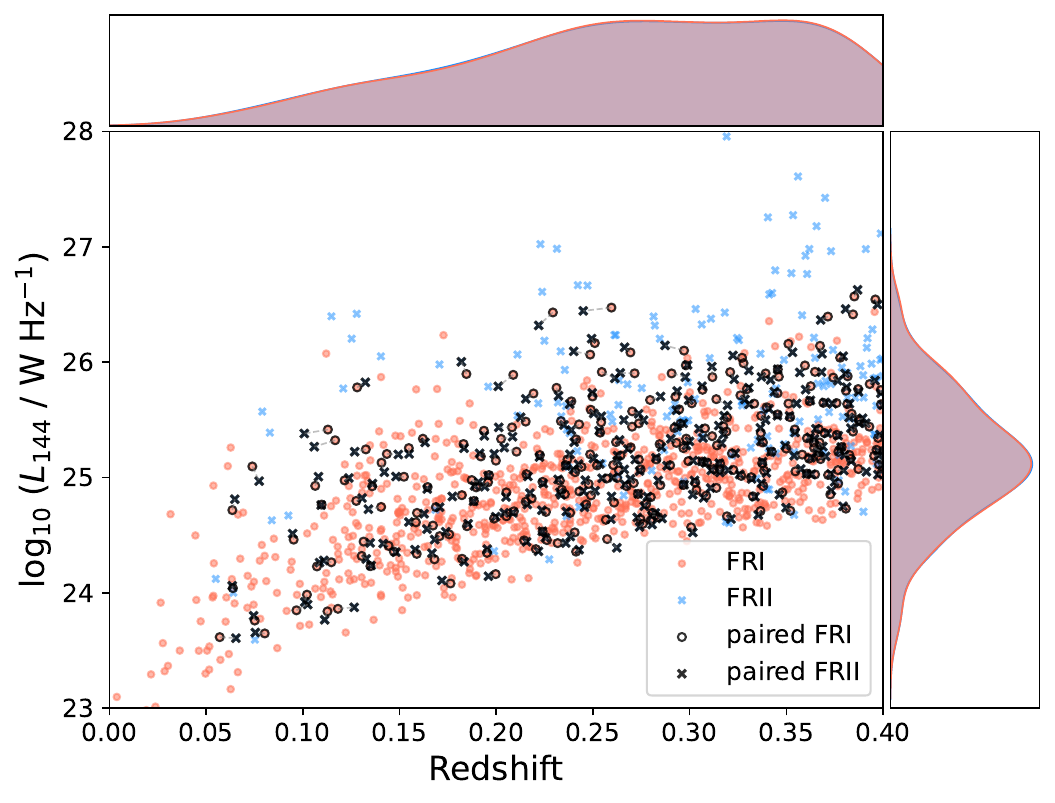}
    \caption{Distribution of radio luminosity at 144 MHz and redshift (\(z\)) for FRIs (red circles) and FRIIs (blue crosses) in the redshift range $0<z<0.4$. The open black circles and black crosses denote the paired FRI and FRII subsamples, respectively. The marginal kernel density estimates (KDEs) along the top and right axes illustrate the one-dimensional distributions of the paired subsamples in redshift and radio luminosity. The paired FRIs and FRIIs are well matched in both properties, enabling unbiased comparisons of their environments.}
    \label{fig:match}

\end{figure}

    To further reduce the biases arising from differing redshift and luminosity distributions between FRI and FRII sources, we constructed a paired sample at $z<0.4$ by identifying FRI-FRII pairs with the closest redshift and luminosity on the $p$-$z$ diagram. This results in 261 pairs, as is illustrated in Fig. \ref{fig:match}. Among them, 177 FRIs and 122 FRIIs have spectroscopic redshifts. This paired sample ensures nearly identical redshift and luminosity distributions for FRIs and FRIIs (Fig. \ref{fig:match}), enabling a cleaner investigation into the intrinsic physical differences between these two classes of radio sources. To ensure that the paired sample provides a fair comparison between FRIs and FRIIs, we tested whether the two subsamples have similar distributions in redshift and radio luminosity, both globally and within the luminosity bins used in Fig. \ref{fig:match}. For the full paired sample ($261$ FRI and FRII pairs), a two-sample Kolmogorov--Smirnov test yields $D=0.019$ with $p>0.99$ for both redshift and luminosity, indicating no evidence of a mismatch between the paired subsamples. We repeated the same test within each luminosity bin and again found no significant differences between the FRI and FRII redshift or luminosity distributions (all $p \ge 0.87$ for redshift and all $p \ge 0.97$ for luminosity). Finally, we assessed the stability of the pairing against finite-sample fluctuations by regenerating the paired sample $2000$ times and repeating the tests; the resulting $p$ value distributions remain consistently high (median $p=0.9997$ for redshift and $p=0.998$ for luminosity). These tests indicate that the pairing procedure is unlikely to introduce a significant systematic bias in redshift or luminosity, enabling a reliable comparison of the environments of FRIs and FRIIs.

    In summary, we define two complementary samples at $z<0.4$ to robustly compare FRIs and FRIIs. To minimise observational biases while ensuring reliable statistics, a volume-limited sample with strict redshift and luminosity cuts are used. This allows us to retain the most luminous FRIIs, which are complemented by a paired sample that matches FRIs and FRIIs in both redshift and luminosity. The paired method ensures that the two populations are directly comparable, which allows for a more reliable investigation into their intrinsic properties and local environmental dependences.

\subsection{Galaxy cluster sample}

    To investigate the cluster-scale environments of FRI and FRII radio galaxies, we adopted the galaxy cluster catalogue compiled by \citet[][hereafter W24]{Wen2024}. The W24 catalogue is constructed based on the DESI Legacy Imaging Surveys \citep{Dey2019AJ....157..168D} and incorporates available spectroscopic redshifts. These spectroscopic redshifts are primarily drawn from the Two Micron All-Sky Survey (2MASS) Redshift Survey \citep{Huchra2012ApJS..199...26H}, the SDSS DR17 \citep{Abdurro'uf2022ApJS..259...35A}, and the early data release of DESI \citep{DESICollaboration2024AJ....168...58D}. Galaxy clusters in W24 are identified by detecting overdensities in the stellar mass distribution of galaxies within redshift slices centred on pre-selected massive galaxies. In total, the catalogue contains 1.58 million galaxy clusters. 

    The W24 catalogue provides a widely used optical cluster sample and shows good consistency with X-ray and Sunyaev--Zeldovich (SZ) selected cluster catalogues. For clusters with halo masses of $M_{500} \geq 0.47 \times 10^{14}~M_\odot$, it recovers over 95\% of SZ-selected clusters from \textit{Planck} \citep{PlanckCollaboration2016A&A...594A..27P} and about 90\% of X-ray selected clusters from ROSAT and eROSITA surveys \citep{Klein2023MNRAS.526.3757K,Bulbul2022A&A...661A..10B,Liu2022A&A...661A...2L}. Within the redshift range $z < 0.4$, approximately 59,000 clusters lie within the LoTSS-DR2 footprint. Among these, 62\% have spectroscopic redshifts, while the remainder have photometric redshifts with a typical uncertainty of 0.013.

    As with any photometric cluster catalogues, W24 is also subject to residual systematics, including incompleteness and contamination, which are not fully quantified. In this work, we assume that these effects primarily impact the absolute number of cluster associations, while having a limited impact on the relative comparison between FRI and FRII populations.

\subsection{Cross-matching FRIs and FRIIs with the galaxy cluster catalogue}

    To systematically characterise the cluster-scale environments of FRI and FRII radio galaxies, we cross-matched our radio galaxy samples with the W24 galaxy cluster catalogue. Following a similar approach to \citet{Pan2025A&A...695A..69P}, an FRI or FRII is considered to be associated with a galaxy cluster if the following two criteria are both satisfied:
    
    \begin{enumerate}
        \item The redshift difference, $\Delta z$, between the FRI or FRII and the galaxy cluster is below 0.01. This threshold is justified by the fact that most of the FRIs, FRIIs, and clusters have accurate spectroscopic redshifts, and the typical uncertainty in photometric redshifts is also of the order of 0.01. To verified that our adopted redshift tolerance (\(\Delta z \leq 0.01\)) does not introduce a significant number of chance associations, we computed the empirical \(\Delta z\) distribution for all radio galaxy-cluster matches. The offsets are extremely small: the median \(\Delta z\) is \(4.3\times10^{-5}\), the 95th percentile is 0.0074, and 86\% (92\%) lie within \(\Delta z\) < 0.003 (\(\Delta z\) < 0.005). Only a select number of matches approach the \(\Delta z\) = 0.01 boundary. This confirms that the adopted tolerance is not driving the match statistics. As FRIs and FRIIs are matched under identical criteria, any residual uncertainty affects both classes equally and cannot generate the differential trends observed in this work.
        \item The projected distance, $\Delta D$, between the FRI or FRII and the cluster centre is required to be smaller than $2 R_{500}$. For this matching radius, we define the association boundary using $R_{500}$ values provided by W24. $R_{500}$ is defined as the radius within which the average density of the cluster is 500 times the critical density of the Universe, while $R_{200}$ follows the same definition but at 200 times the critical density. Typically, $R_{200}$ is approximately 1.5 to 2 times $R_{500}$ and is often adopted as the boundary of a galaxy cluster \citep[see e.g.][]{Morandi2015MNRAS.450.2261M,Mirakhor2020MNRAS.497.3204M}. To conservatively capture all plausible associations between radio galaxies and clusters, we adopted a matching radius of $2 R_{500}$. We note that the $R_{500}$ values provided in W24 are derived from a photometric cluster-finding algorithm and therefore may carry redshift-dependent systematics. In this work, $R_{500}$ is used only as an operational, catalogue-consistent aperture for defining cluster association, rather than as a mass proxy. We do not adopt a radio galaxy-based radius because our aim is to probe the cluster-scale environment, and radio and host sizes are intrinsic source properties that can differ systematically between FRIs and FRIIs. Because our analysis is differential (FRIs versus FRIIs) and restricted to the same redshift range (and to matched redshift distributions in the paired sample), any residual systematic uncertainty in $R_{500}$ affects both classes similarly and does not influence our comparative conclusions.
    \end{enumerate}
    
    To test whether our matching procedure introduces any systematic or class-dependent bias between FRIs and FRIIs, we constructed 50 mock catalogues for each class. The mock samples contain the same numbers of sources as the volume-limited sample (\(N=767\) for FRIs and \(N=369\) for FRIIs), while preserving the redshift and \(L_{144}\) distributions of each class. We then randomised the sky positions within the region \(180^\circ < {\rm RA} < 210^\circ\) and \(30^\circ < {\rm Dec} < 60^\circ\), covered by the W24 cluster catalogue, and applied exactly the same matching criteria as for the real samples; namely, \(\Delta z < 0.01\) and \(\Delta D < 2R_{500}\).
    
    Only a very small fraction of the mock sources are matched to W24 clusters. The mean mock match fractions are nearly identical for the two classes, with \(\langle f_{\rm FRI} \rangle = 0.021\) and \(\langle f_{\rm FRII} \rangle = 0.020\),
    which indicates that the matching procedure is unlikely to introduce a preferential association of either FRIs or FRIIs with clusters.
    Since the mock tests were performed in a relatively compact sky region, corresponding to a higher surface density of random sources than in the real samples, these values should be regarded as conservative upper limits on the chance-coincidence fraction. The substantially higher match fractions measured for the real radio galaxies are therefore unlikely to be driven by random associations.

\section{Results} \label{sec3}

    We now investigate the environments of FRI and FRII radio galaxies using their associations with galaxy clusters from the W24 catalogue. Our analysis begins by quantifying the fraction of FRIs and FRIIs matched to clusters, and how this fraction varies with radio luminosity, stellar mass, and host galaxy brightness. We then examine whether the spatial positions of these sources within clusters and the properties of the clusters themselves differ between the two classes. Finally, we assess the likelihood that FRIs and FRIIs serve as brightest cluster galaxies (BCGs), and how this role correlates with their host galaxy properties.

\subsection{The cluster match fractions and their dependence on source properties} \label{sec:result1}

    We first quantified the fraction of FRIs and FRIIs that are associated with galaxy clusters in the W24 catalogue. In total, 373 FRIs and 113 FRIIs in the volume-limited sample, and 119 FRIs and 85 FRIIs in the paired sample, have a cluster counterpart. Table \ref{tab:tab1} summarises the cluster match fractions within $2R_{500}$. All error bars throughout this work are 68\% Jeffreys binomial confidence intervals \citep{Jeffreys1946RSPSA.186..453J}. For an observed cluster match fraction, $f=k/n$, we adopt the Jeffreys posterior
    \begin{equation}
        p(f\mid k,n)\propto \mathrm{Beta}(k+1/2,\,n-k+1/2)\,,
    \end{equation}
    where $k$ is the number of galaxies that are matched to clusters, and $n$ is the total number of galaxies. 
    In both samples, approximately half of the FRIs but only about one third of the FRIIs are found in clusters.

    To evaluate the statistical significance of this overall difference, we followed the Bayesian framework of \citet{Rickel2025ApJ...983..138R}. 
    Using the Jeffreys posterior distributions of the observed match fractions, we computed the posterior probability that the FRI cluster match fraction exceeds that of FRIIs. 
    For the full sample, this probability exceeds 99.9\%, corresponding to a one-tailed Gaussian significance above $3\sigma$.
    This indicates a statistically significant difference between the cluster match fractions of FRIs and FRIIs within the adopted W24-based matching framework.
    The list of ten examples of FRIs and FRIIs and their associated clusters is provided in Table \ref{tab:10_fri_frii_clusters}.

\begin{table}[]
\centering
\caption{Match fraction with W24 of FRI/FRII within 2$R_{500}$.}
\label{tab:tab1}
\resizebox{0.4\textwidth}{!}{%
\renewcommand{\arraystretch}{1.4}

\begin{tabular}{ccc}
\hline
\multicolumn{1}{l}{} & volume-limit sample & paired sample                 \\ \hline
FRI                  & $48.6^{+1.8}_{-1.8}\%$         & $45.6^{+3.1}_{-3.1}\%$                    \\
FRII                 & $30.6^{+2.5}_{-2.3}\%$        & $32.6^{+3.0}_{-2.8}\%$ \\ \hline
\end{tabular}
}
\end{table}

\begin{figure}[h]
    \centering
    \includegraphics[width=1\linewidth]{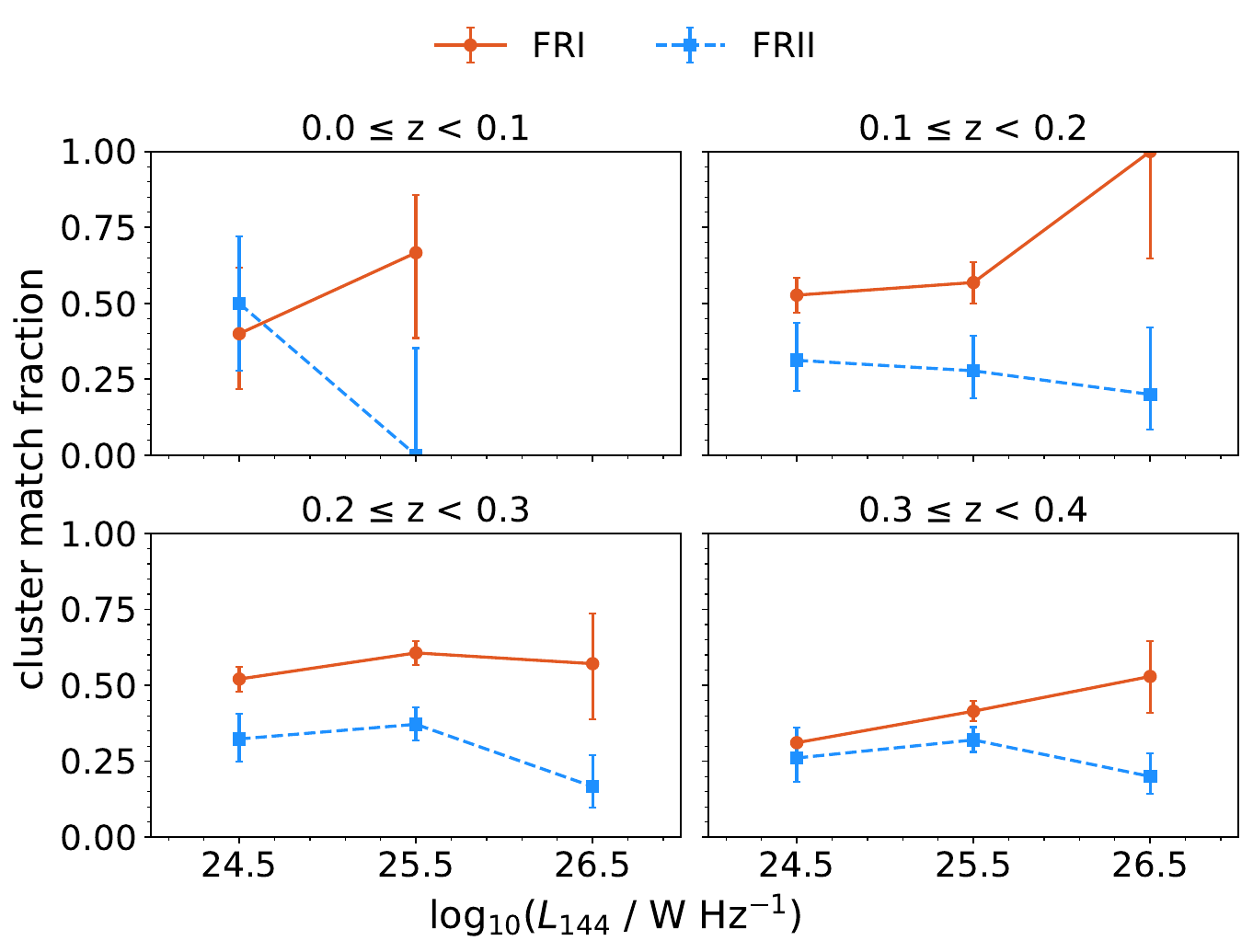}
    \caption{Cluster match fraction as a function of radio luminosity in four redshift bins in the volume-limited sample. Error bars show 68\% Jeffreys intervals. The FRI-FRII separation is present in every redshift interval, demonstrating that the global luminosity-environment relation is not driven by redshift-dependent incompleteness in W24.}
    \label{fig:zslice}
\end{figure}

    We next examined whether this difference depends on redshift. Figure~\ref{fig:zslice} shows the cluster match fraction as a function of 144\,MHz luminosity in four redshift bins. Within each bin, FRIs exhibit a higher cluster match fraction than FRIIs, and neither population shows a monotonic dependence on redshift. Hence, the difference of match fractions between FRI and FRII does not depend on redshift.

\begin{figure}[h]
    \centering

    \begin{subfigure}[b]{0.48\textwidth}
        \centering
        \includegraphics[width=\textwidth]{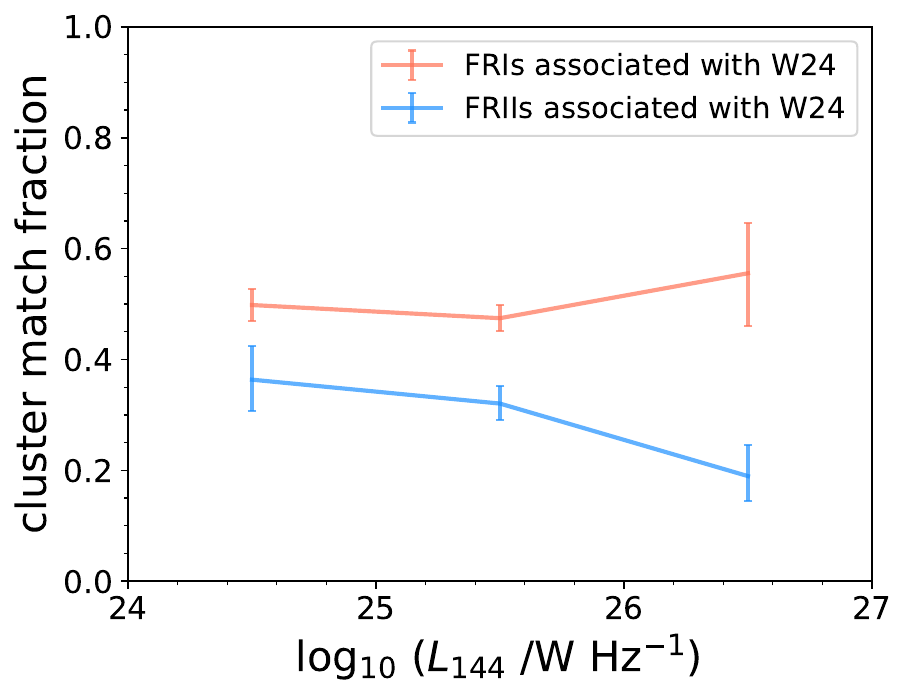}
        \subcaption{}
    \end{subfigure}
    \hfill
    \begin{subfigure}[b]{0.48\textwidth}
        \centering
        \includegraphics[width=\textwidth]{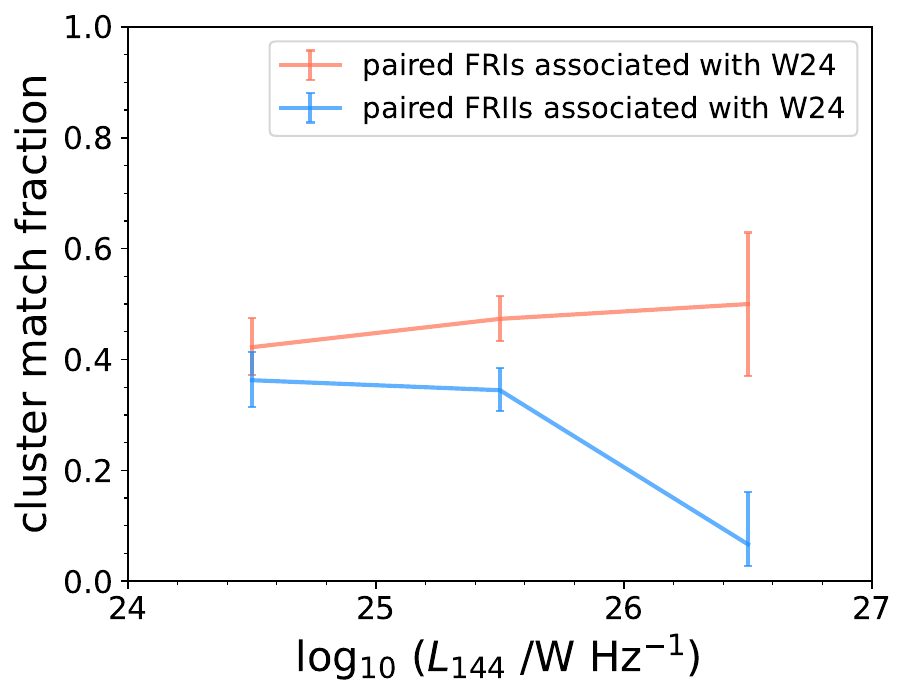}
        \subcaption{}
    \end{subfigure}

    \caption{Comparison of the cluster match fraction and radio luminosity for FRIs and FRIIs within $2R_{500}$:  volume-limited sample (a), paired sample (b).}
    \label{fig:frac_vs_lumi}
\end{figure}

    We present the match fraction as a function of luminosity for both the volume-limited and paired samples in Fig.~\ref{fig:frac_vs_lumi}. 
    In the volume-limited sample, FRIs consistently show higher cluster match fractions than FRIIs at all luminosities. 
    In the paired sample, this difference decreases below $10^{26}\,\mathrm{W\,Hz^{-1}}$, but luminous FRIIs ($>10^{26}\,\mathrm{W\,Hz^{-1}}$) still show the lowest match fractions. 
    The posterior probability that luminous FRIs have a higher match fraction than luminous FRIIs is 99.96\% in the volume-limited sample and 99.7\% in the paired sample, corresponding to one-tailed Gaussian significances of 3.35$\sigma$ and 2.72$\sigma$, respectively. 
    Therefore, the difference remains statistically significant in the high-luminosity regime within the adopted W24-based matching framework.

\begin{figure}[h]
    \centering

    \begin{subfigure}[b]{0.5\textwidth}
        \centering
        \includegraphics[width=\textwidth]{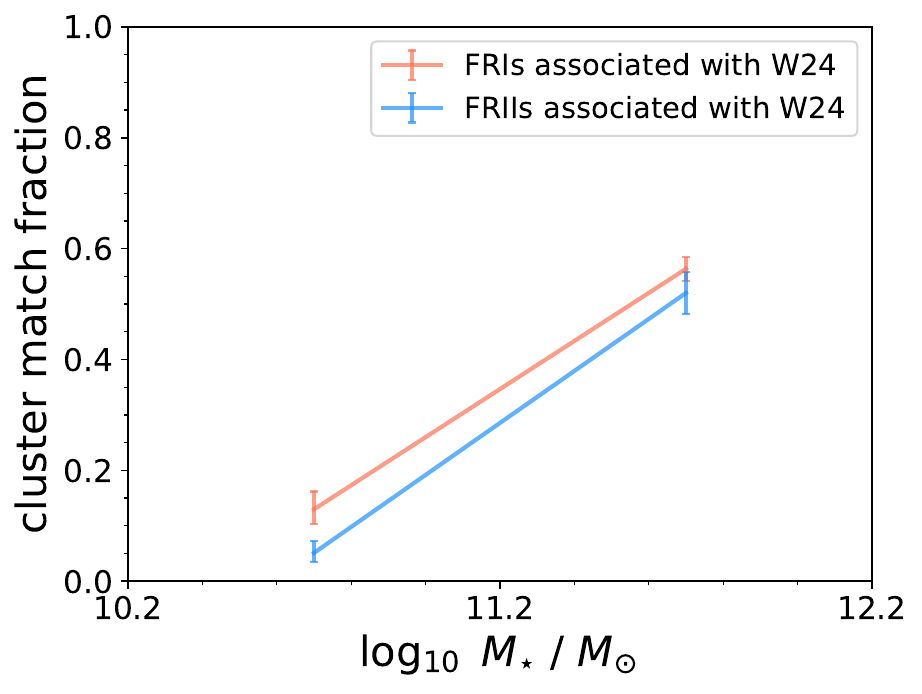}
        \subcaption[]{}
        \end{subfigure}

    \begin{subfigure}[b]{0.5\textwidth}
        \centering
        \includegraphics[width=\textwidth]{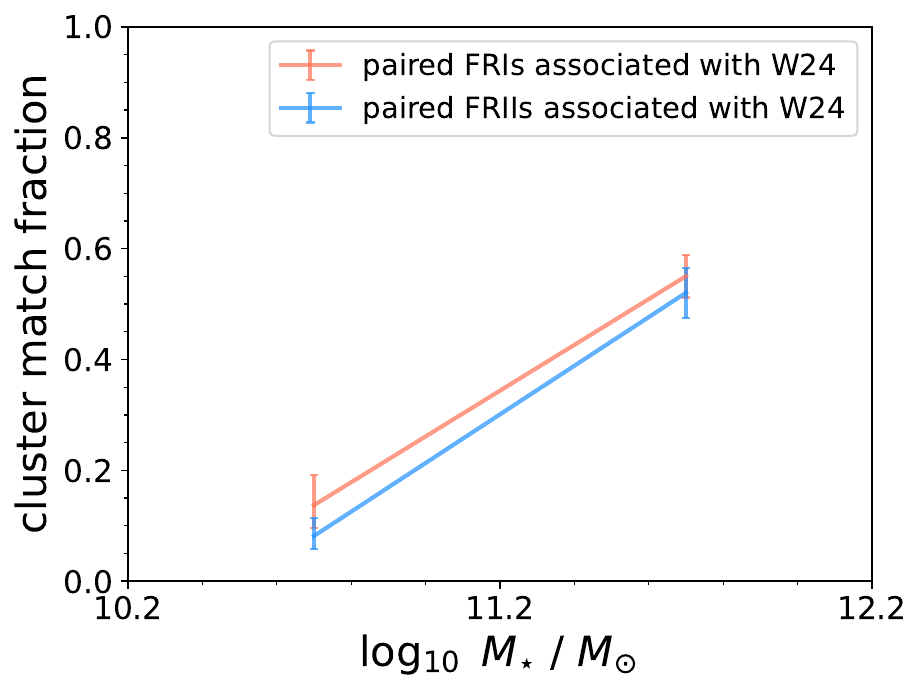}
        \subcaption[]{}
    \end{subfigure}

    \caption{Relationship between the cluster match fraction and stellar mass for FRIs and FRIIs within $2R_{500}$ in volume limit sample (top panel) and paired sample (bottom panel). }
    \label{fig:frac_vs_mass}

\end{figure}

    We further investigated whether the match fraction is related to the stellar mass of the host galaxies (given by the catalogue of \citealt{clews2025MNRAS.tmp..936C}), as shown in Fig. \ref{fig:frac_vs_mass}. For both FRIs and FRIIs, sources with higher stellar masses tend to have higher match fractions. This trend is consistent across both samples. In both cases, when the stellar mass exceeds $10^{11} M_\odot$, the match fractions of FRIs and FRIIs reach up to approximately 50\%. Specifically, in the volume-limited sample, the fractions are \(56.4^{+2.1}_{-2.1}\%\) for FRIs and \(52.0^{+3.8}_{-3.7}\%\) for FRIIs, while in the paired sample they are \(55.0^{+3.8}_{-3.8}\%\) and \(52.0^{+4.5}_{-4.5}\%\), respectively.

\begin{figure}[h]
    \centering

    \begin{subfigure}[b]{0.5\textwidth}
        \centering
        \includegraphics[width=\textwidth]{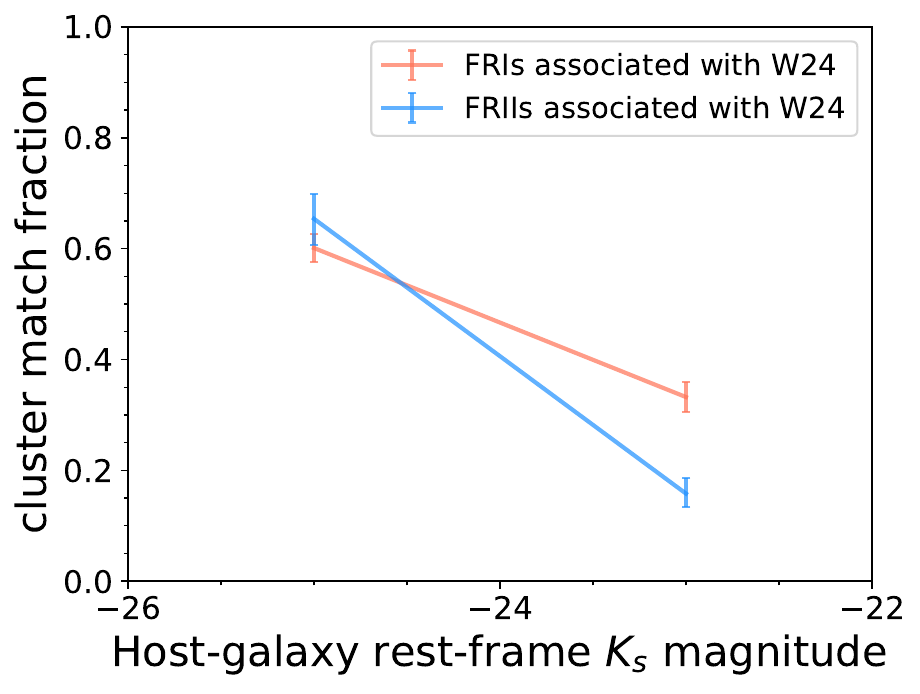}
        \subcaption[]{}
        \end{subfigure}

    \begin{subfigure}[b]{0.5\textwidth}
        \centering
        \includegraphics[width=\textwidth]{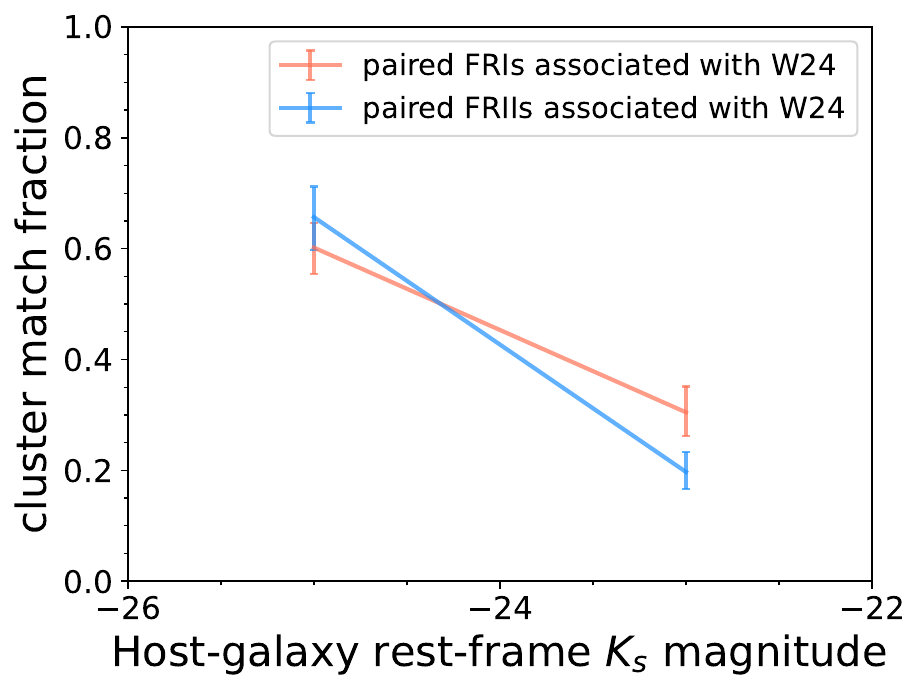}
        \subcaption[]{}
    \end{subfigure}

    \caption{Relationship between the cluster match fraction and the host-galaxy $K_{s}$ magnitude for FRIs and FRIIs within $2R_{500}$ in the volume-limited sample (top panel) and paired sample (bottom panel). }
    \label{fig:frac_vs_ks}

\end{figure}

   We then examined whether the match fraction correlates with the host galaxy’s $K_{s}$-band magnitude, which is commonly used as a proxy for stellar mass \citep{Konishi2011PASJ...63S.363K,Mingo2022MNRAS.511.3250M,clews2025MNRAS.tmp..936C}, as shown in Fig. \ref{fig:frac_vs_ks}. For both FRIs and FRIIs, sources with brighter host galaxies tend to have higher match fractions. This trend is consistent across both samples. For the paired sample, the match fractions of FRIs and FRIIs are more similar across different host galaxy brightness levels compared to those in the volume-limited sample.

\subsection{Relations between FRI and FRII properties and the properties of their associated clusters}

\begin{figure}[h]
    \centering

    \begin{subfigure}[b]{0.5\textwidth}
        \centering
        \includegraphics[width=\textwidth]{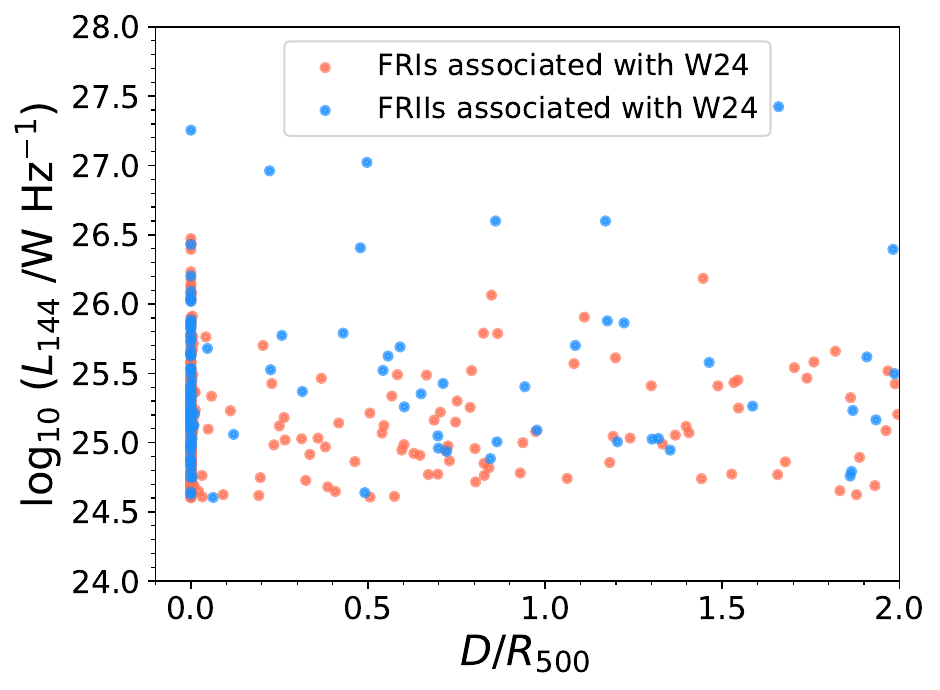}
        \subcaption[]{}
        \end{subfigure}

    \begin{subfigure}[b]{0.5\textwidth}
        \centering
        \includegraphics[width=\textwidth]{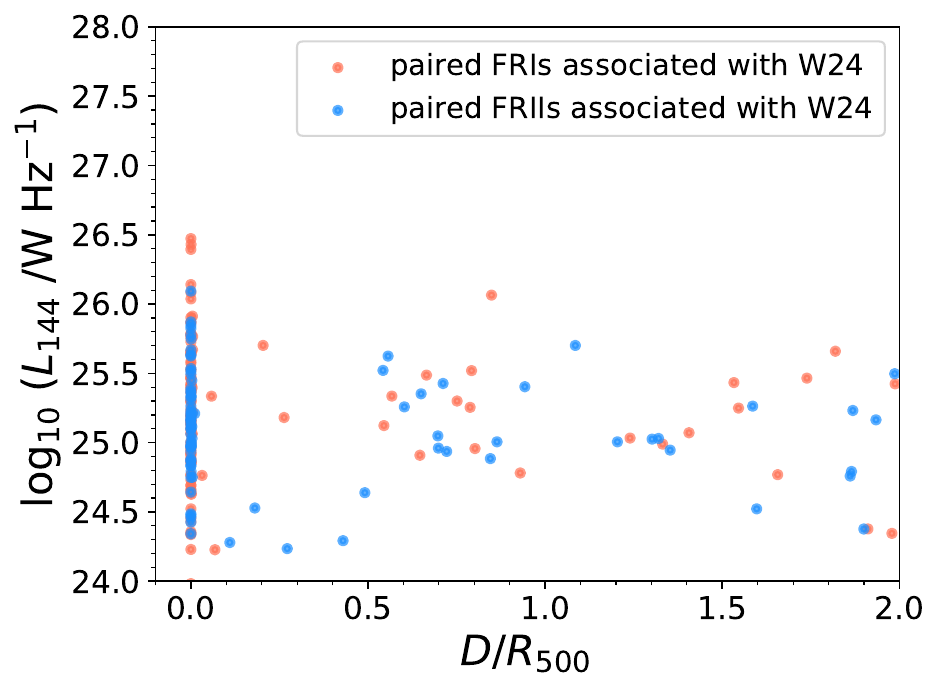}
        \subcaption[]{}
    \end{subfigure}

    \caption{Relationship between the luminosity and distance to cluster centre for FRIs and FRIIs within $2R_{500}$ in the volume-limited sample (top panel) and paired sample (bottom panel). }
    \label{fig:lumi_vs_dis}

\end{figure}

\begin{figure}[h]
    \centering

    \begin{subfigure}[b]{0.5\textwidth}
        \centering
        \includegraphics[width=\textwidth]{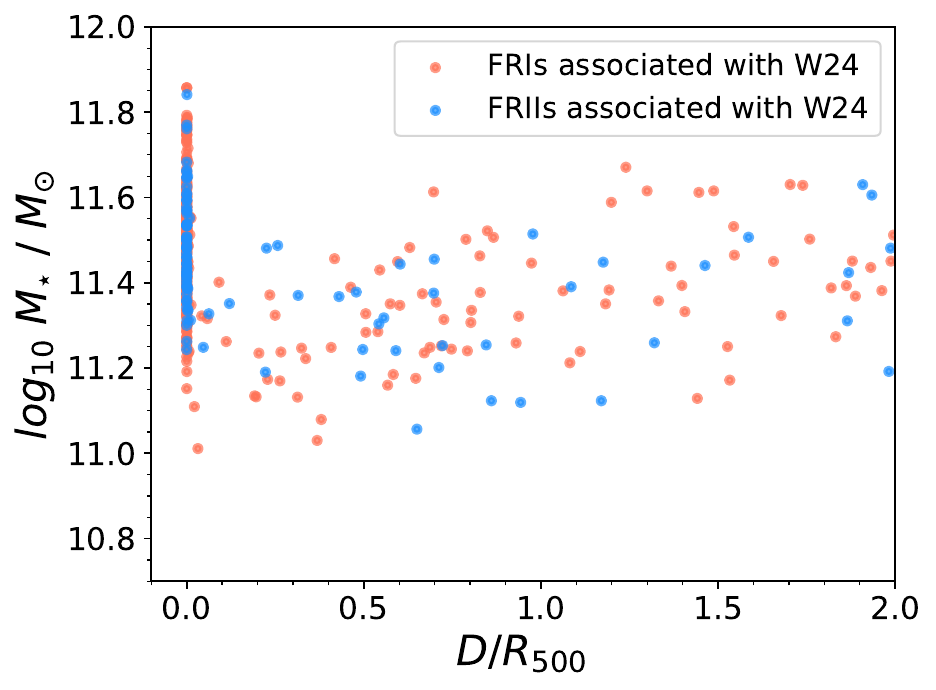}
        \subcaption[]{}
        \end{subfigure}

    \begin{subfigure}[b]{0.5\textwidth}
        \centering
        \includegraphics[width=\textwidth]{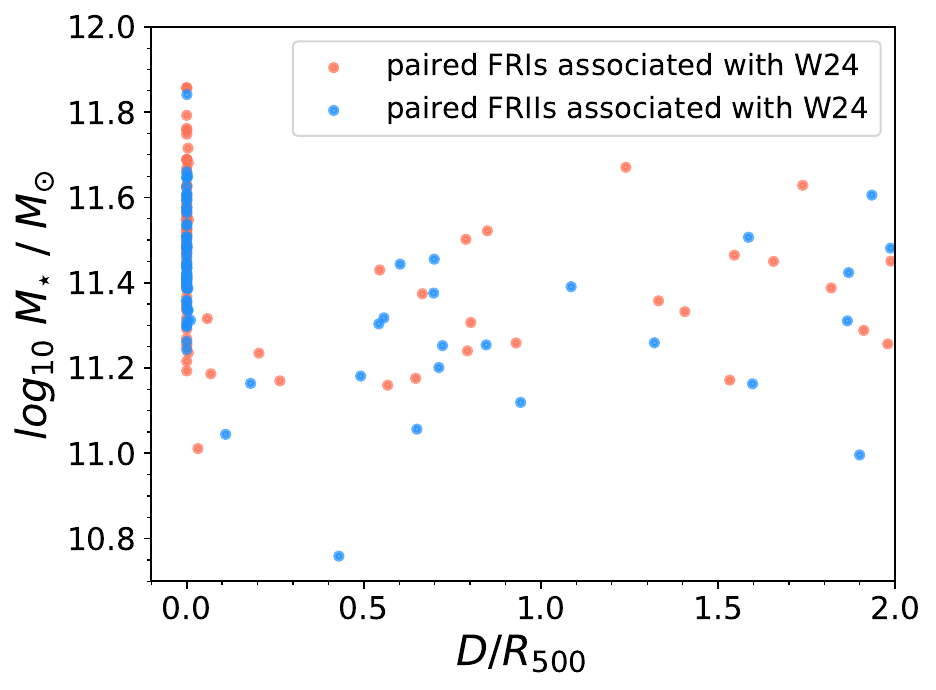}
        \subcaption[]{}
    \end{subfigure}

    \caption{Relationship between the mass and distance to cluster centre for FRIs and FRIIs within $2R_{500}$ in volume limit sample (top panel) and paired sample (bottom panel). }
    \label{fig:mass_vs_dis}

\end{figure}

    We investigated possible dependences of the properties of FRIs and FRIIs on their positions within galaxy clusters. Figure~\ref{fig:lumi_vs_dis} shows radio luminosity as a function of projected distance from the cluster centre (normalised by $R_{500}$) for FRIs and FRIIs associated with clusters. The top and bottom panels correspond to the volume-limited and paired samples, respectively. Similarly, Fig.~\ref{fig:mass_vs_dis} shows stellar mass as a function of projected cluster-centric distance for the same sources. In both Figs.~\ref{fig:lumi_vs_dis} and \ref{fig:mass_vs_dis}, several sources lie very close to the cluster centres ($D < 10$ kpc). We identify these sources as the BCGs, since their projected distances from the cluster centres are smaller than the typical size of a BCG (approximately 20 kpc), and no other galaxies are detected within 100 kpc of the cluster centres. However, we find no significant correlation between radio luminosity or stellar mass and the projected distance to the cluster centre.

\begin{figure}[h]
    \centering

    \begin{subfigure}[b]{0.5\textwidth}
        \centering
        \includegraphics[width=\textwidth]{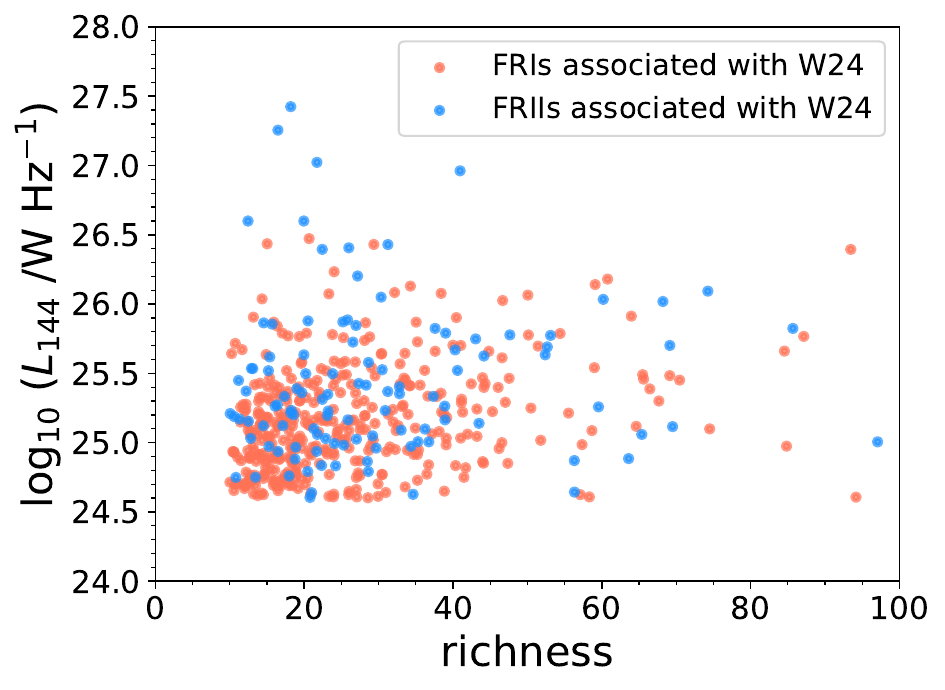}
        \subcaption[]{}
        \end{subfigure}

    \begin{subfigure}[b]{0.5\textwidth}
        \centering
        \includegraphics[width=\textwidth]{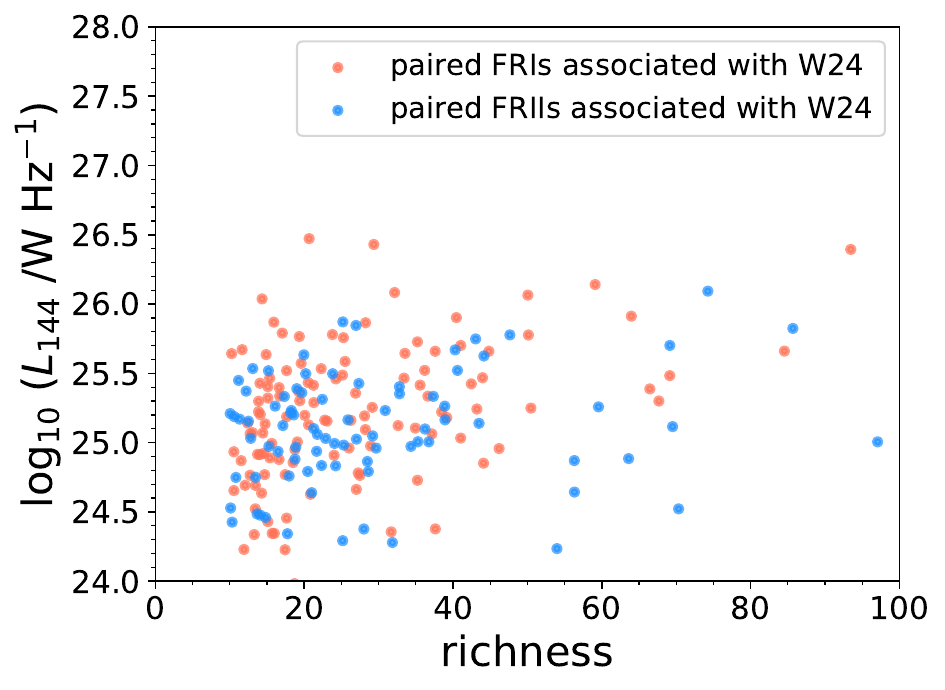}
        \subcaption[]{}
    \end{subfigure}

    \caption{Relationship between the radio luminosity and richness of the matched cluster for FRIs and FRIIs within $2R_{500}$ in volume limit sample (top panel) and paired sample (bottom panel). }
    \label{fig:lumi_vs_rich}

\end{figure}

\begin{figure}[h]
    \centering

    \begin{subfigure}[b]{0.5\textwidth}
        \centering
        \includegraphics[width=\textwidth]{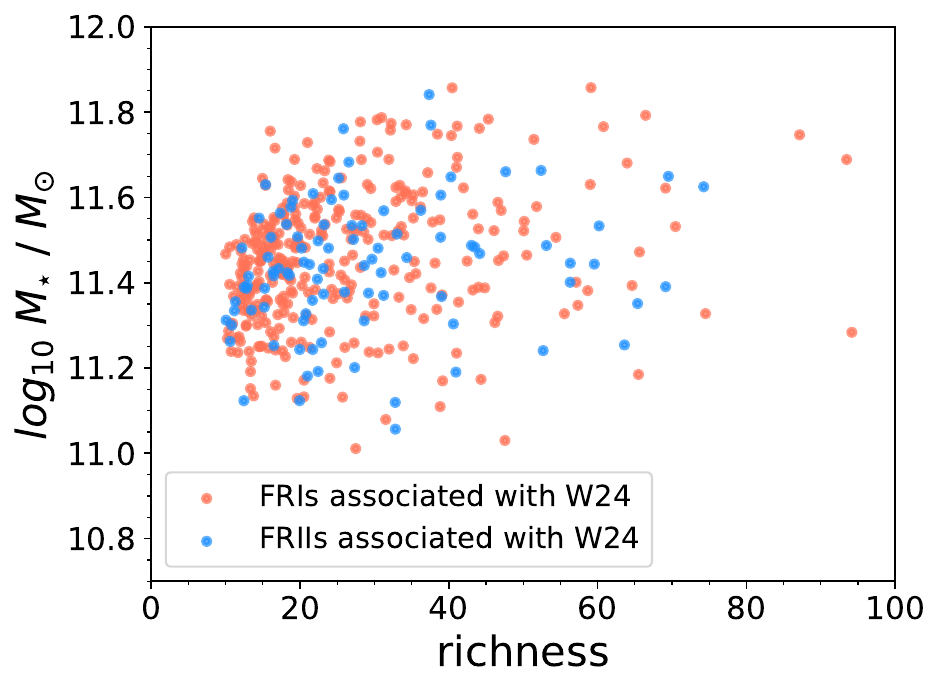}
        \subcaption[]{}
        \end{subfigure}

    \begin{subfigure}[b]{0.5\textwidth}
        \centering
        \includegraphics[width=\textwidth]{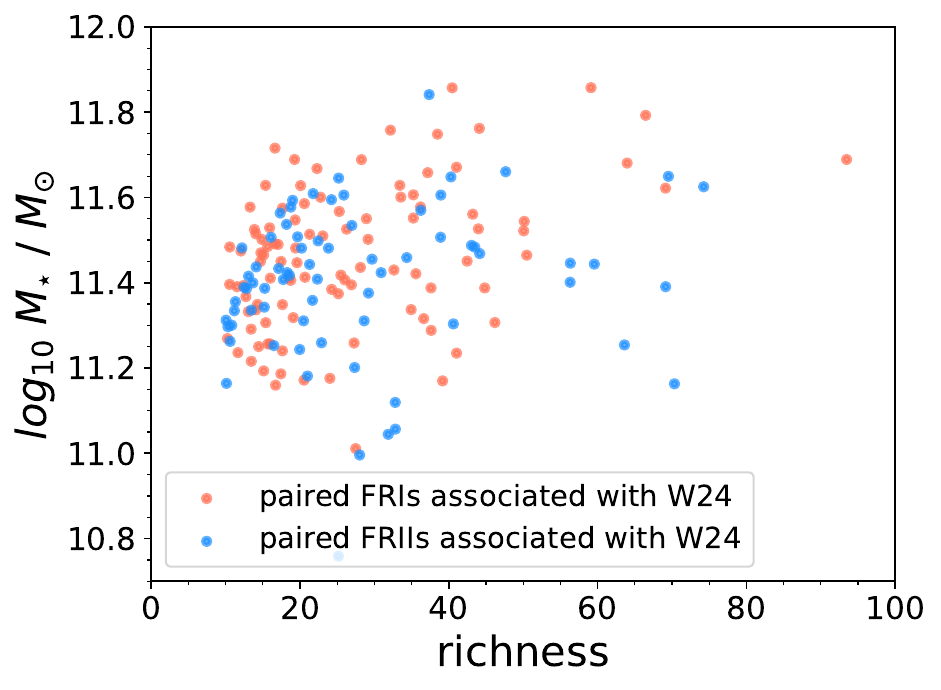}
        \subcaption[]{}
    \end{subfigure}

    \caption{Relationship between the stellar mass and richness of the matched cluster for FRIs and FRIIs within $2R_{500}$ in volume limit sample (top panel) and paired sample (bottom panel). }
    \label{fig:mass_vs_rich}

\end{figure}

    We also investigated whether these radio galaxy properties correlate with the richness of the host clusters. Figure~\ref{fig:lumi_vs_rich} shows the radio luminosity of the galaxies versus the richness of the host clusters, and Fig.~\ref{fig:mass_vs_rich} shows the stellar mass versus the cluster richness. For both volume-limited and paired samples, most FRIs and FRIIs are found in clusters with richness between 10 and 30. The distributions for the FRIs and FRIIs are broadly similar, and no significant trends or differences are observed.

    Overall, our results indicate no significant dependence of FRI or FRII properties on their positions within clusters or on global cluster characteristics such as richness. This conclusion holds for both volume-limited and paired samples.

\subsection{Radial distributions of FRI and FRII surface densities in clusters}

\begin{figure}[h]
    \centering

    \begin{subfigure}[b]{0.5\textwidth}
        \centering
        \includegraphics[width=\textwidth]{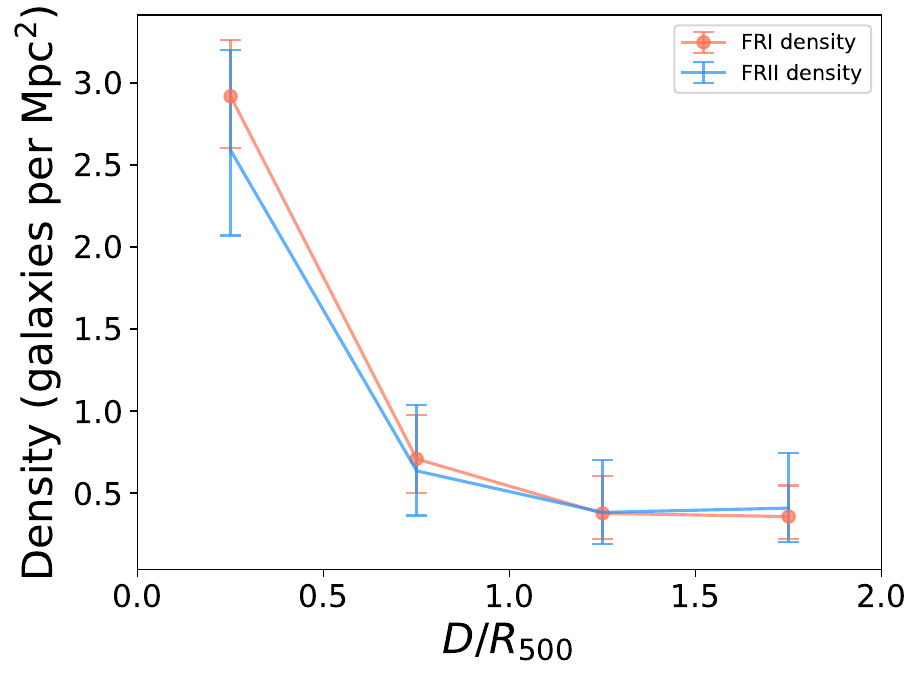}
        \subcaption[]{}
        \end{subfigure}

    \begin{subfigure}[b]{0.5\textwidth}
        \centering
        \includegraphics[width=\textwidth]{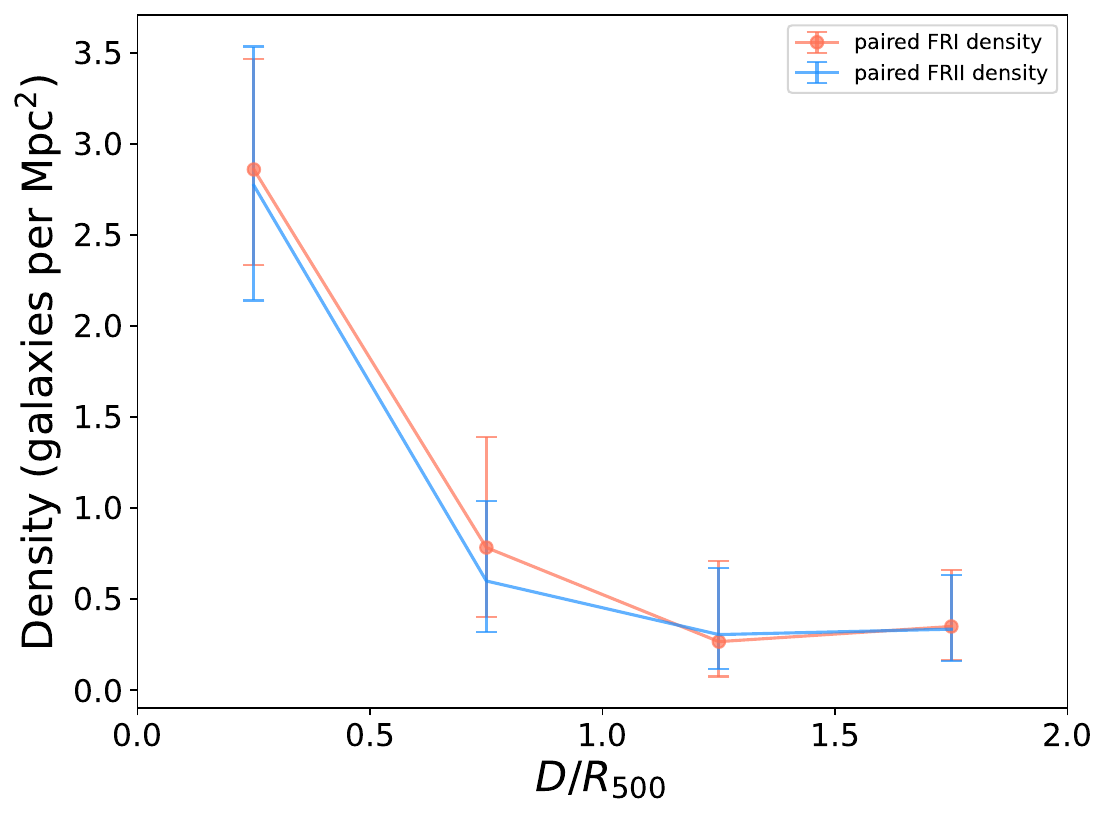}
        \subcaption[]{}
    \end{subfigure}

    \caption{Relationship between the densities and distance to cluster centres for FRIs and FRIIs within $2 \times R_{500}$ in volume limit sample (top panel) and paired sample (bottom panel). }
    \label{fig:density}

\end{figure}

    Previous studies have shown that both FRI and FRII can reside near the centres of galaxy clusters and are often BCGs \citep{pedlar1990MNRAS.246..477P}. At lower redshifts, FRIs tend to be more centrally concentrated than FRIIs within clusters \citep{Vardoulaki2021A&A...648A.102V}. To further investigate this trend, we computed the projected surface number density profiles of FRIs and FRIIs within their associated galaxy clusters. The density in each radial bin was calculated from the cluster centre as the number of sources within a given annulus divided by the area of that annulus and is expressed per square megaparsec. The resulting radial distributions are shown in Fig.~\ref{fig:density}.

    Overall, regardless of the sample considered, the highest densities for both FRIs and FRIIs are at around $0.5 \times R_{500}$, with average source densities exceeding 2 Mpc$^{-2}$. However, when the distance from the cluster centre exceeds $R_{500}$, the densities of both FRIs and FRIIs decrease and gradually stabilise. We do not find any significant difference in the density distributions between FRIs and FRIIs.

\subsection{Roles of FRI and FRII in clusters}
    
    Next, we examined whether FRIs and FRIIs are equally likely to serve as the BCGs, defined as radio sources located within 10 kpc of the cluster centre. In the volume-limited sample, $74.8^{+2.2}_{-2.3}\%$ of FRIs are identified as BCGs, compared to $61.9^{+4.4}_{-4.6}\%$ of FRIIs. The same trend is seen in the paired sample, where the BCG fractions are $78.1^{+3.5}_{-4.0}\%$ for FRIs and $65.9^{+4.9}_{-5.3}\%$ for FRIIs. The Bayesian posterior probability that FRIs are more likely than FRIIs to be BCGs corresponds to a one-tailed significance of 2.6$\sigma$ for the volume-limited sample and 1.9$\sigma$ for the paired sample, indicating modest but consistent differences across both samples. These results suggest that FRIs are more likely than FRIIs to occupy the central positions of galaxy clusters. We further compare the physical properties of BCG-hosted FRIs and FRIIs, as summarised in Table \ref{tab:median}. Table \ref{tab:median} provides the median values for key physical parameters: radio luminosity ($\log_{10}(L_{144})$), host galaxy stellar mass ($\log_{10}(M_\star/M_\odot)$), radio source physical size, cluster richness, cluster halo mass ($M_{500}$), and the stellar mass of the second brightest cluster galaxy ($\log_{10}(M_{\star,2}/M_\odot)$). Our analysis shows that there are no substantial differences between the two classes once they are both BCGs.

\begin{table*}[]
    \centering
    \small 
    \caption{Median properties of FRI and FRII BCGs in the volume-limited and paired samples.\label{tab:median} }
    \begin{tabular}{llccccccc}
        \hline
        Sample & Class 
        & $\log_{10}(L_{144}~[\mathrm{W\,Hz}^{-1}])$ 
        & $\log_{10}(M_\star/M_\odot)$
        & Size [kpc] 
        & Cluster Richness 
        & $M_{500}~[10^{14}~M_\odot]$ 
        & $\log_{10}(M_{\star,2}/M_\odot)$ \\
        \hline
        \multirow{2}{*}{Volume-limited} 
               & FRI  & $25.16 \pm 0.03$ & $11.60 \pm 0.01$ & $292.5 \pm 6.7$  & $19.6 \pm 0.8$  & $0.90 \pm 0.04$ & $10.19 \pm 0.01$ \\
               & FRII & $25.34 \pm 0.07$ & $11.59 \pm 0.04$ & $294.3 \pm 25.0$ & $24.2 \pm 1.6$  & $1.10 \pm 0.07$ & $10.16 \pm 0.02$ \\
        \multirow{2}{*}{Paired}     
               & FRI  & $25.19 \pm 0.06$ & $11.56 \pm 0.03$ & $265.8 \pm 8.7$  & $20.1 \pm 1.5$  & $0.92 \pm 0.06$ & $10.19 \pm 0.02$ \\
               & FRII & $25.16 \pm 0.06$ & $11.56 \pm 0.02$ & $309.0 \pm 27.9$ & $21.7 \pm 1.9$  & $0.99 \pm 0.08$ & $10.16 \pm 0.02$ \\
        \hline
     
    \end{tabular}
\end{table*}

\begin{figure}[h]
    \centering

    \begin{subfigure}[b]{0.5\textwidth}
        \centering
        \includegraphics[width=\textwidth]{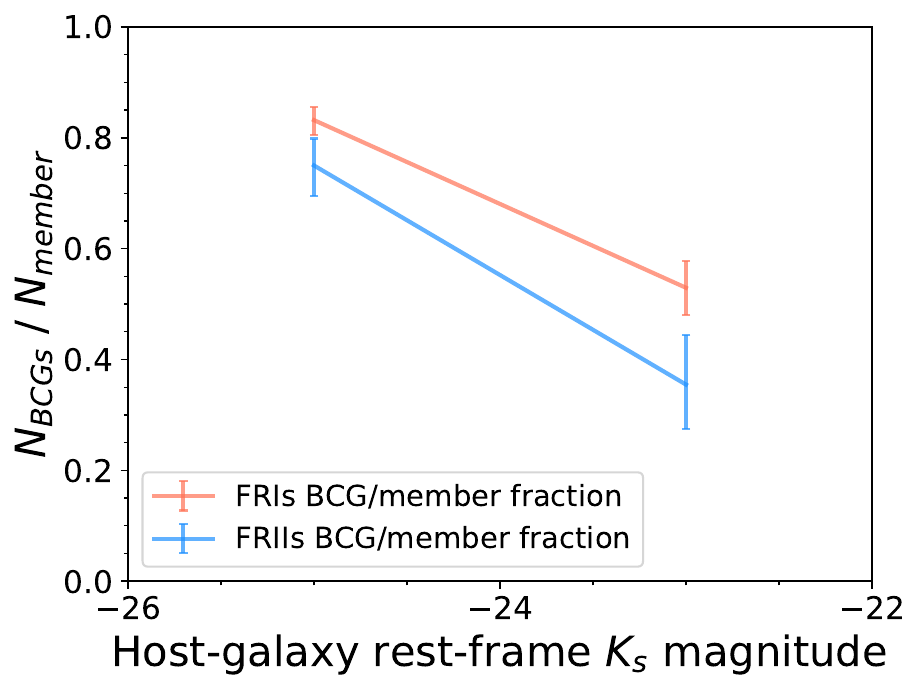}
        \subcaption[]{}
        \end{subfigure}

    \begin{subfigure}[b]{0.5\textwidth}
        \centering
        \includegraphics[width=\textwidth]{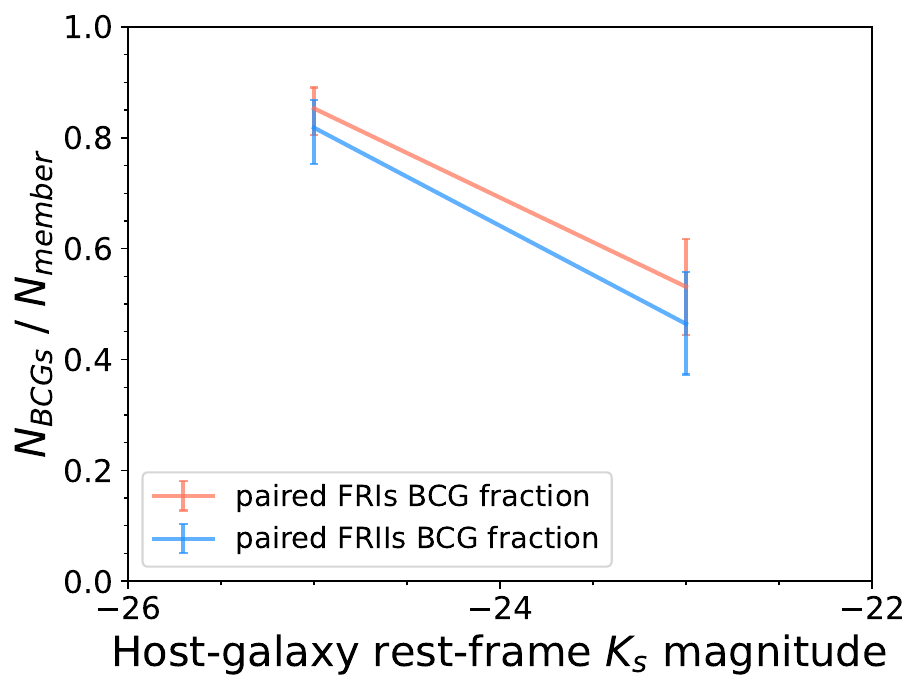}
        \subcaption[]{}
    \end{subfigure}

    \caption{Relationship between the BCG fraction and the host-galaxy $K_{s}$ magnitude for FRIs and FRIIs within 2$R_{500}$. The BCG fraction is defined as the fraction of a given population that are BCGs, compared to those that are not BCGs but are still cluster members. The top panel shows the volume-limited sample, while the bottom panel shows the paired sample.}
    \label{fig:bcgfrac_vs_ks}

\end{figure}

    Finally, we examined how the likelihood of a radio galaxy being the BCG varies with the host galaxy’s rest-frame $K_s$-band magnitude, as shown in Fig.~\ref{fig:bcgfrac_vs_ks}. For both FRIs and FRIIs, the BCG fraction decreases towards fainter host galaxies. The BCG fraction is defined as the fraction of a given population that are BCGs, compared to those that are not BCGs but are still cluster members. In the volume-limited sample, the posterior probability that FRIs have a higher BCG fraction than FRIIs is 93.2\% at $K_s = -25$ (equivalent to 1.5$\sigma$ one-tailed significance), and increases to 95.6\% at $K_s = -23$ (1.7$\sigma$). For the paired sample, the corresponding probabilities are 69.1\% and 69.9\%, or approximately 0.5$\sigma$ in both bins. While FRIs consistently show a slightly higher BCG fraction than FRIIs, the difference remains modest and roughly constant across all bins. This trend is observed in both the volume-limited and paired samples.

\section{Discussion} \label{sec4}

    One of the key results of our study is that FRI galaxies are more likely to reside in clusters than FRIIs. This difference is especially pronounced among high-radio-luminosity FRIIs (see Fig. \ref{fig:frac_vs_lumi}), where only about \(19.0^{+5.7}_{-4.6}\%\) in the volume-limit sample (\(6.7^{+9.5}_{-3.9}\%\) for the paired sample) are found to be cluster members. In other words, at low redshift, powerful FRIIs are more commonly located in the field or in poor group environments, whereas a substantial fraction of FRIs inhabit galaxy clusters. This trend is consistent with the jet disruption model, which proposes that in a dense intracluster medium, surrounding gas can decelerate even relatively powerful jets through drag and turbulence, slowing them to subsonic speeds. This deceleration results in the formation of plume-like structures rather than narrow, collimated jets ending in bright hotspots, thereby favouring an FRI morphology \citep[e.g.][]{Bicknell1994ApJ...422..542B,Laing2002}. In contrast, FRII jets tend to remain well collimated over large scales and are therefore more frequently found in less dense environments. Among the FRII-Lows, the cluster association fraction remains lower than that of FRI sources at comparable luminosities, although the difference is less pronounced. This may also reflect the jet disruption framework: FRII-Lows might inhabit lower-density inner environments, allowing their weaker jets to remain stable and maintain an FRII morphology.

    We also find that the cluster association fraction increases with host stellar mass and $K_s$-band brightness for both FRIs and FRIIs. This is a general result that holds for the whole galaxy population, independent of radio emission: more massive or optically luminous galaxies are preferentially found in denser environments \citep{Muldrew2012MNRAS.419.2670M,Vardoulaki2021A&A...648A.102V}. In our case, FRIs and FRIIs both follow this well-established trend, and we do not find significant differences between the two radio morphological classes. Therefore, the lower cluster match fraction of FRIIs compared to FRIs cannot be attributed to host stellar mass.

    To further assess environmental influence, we compared the richness of the associated clusters for FRIs and FRIIs. We found no significant differences between the two classes in either parameter across both samples. This result contrasts with some earlier studies suggesting that FRIs are more likely to reside in richer clusters \citep[e.g.][]{Wing2011AJ....141...88W,Gendre2013MNRAS.430.3086G}, but those studies typically focused on high-luminosity FRIIs and lacked luminosity control. Our findings suggest that, when matched in radio luminosity, the cluster-scale environment is not the primary driver of FR morphology. The cluster richness values in W24 are algorithm-dependent quantities that may carry redshift-dependent systematics and are not accompanied by formal uncertainty estimates. In this paper, we therefore employ richness only as a relative indicator of the local galaxy density, rather than as an absolute mass proxy. Because our conclusions are based on a differential comparison between FRIs and FRIIs within the same redshift range (and within the redshift-matched paired sample), any residual systematic uncertainty in $R_{500}$ should largely cancel out and does not change our comparative conclusions.

    Moreover, despite the marked difference in overall cluster match fractions, the spatial distributions of FRIs and FRIIs within clusters are similar. Both classes show peak source densities around $0.5 \times R_{500}$, with declining densities beyond $R_{500}$. This consistency suggests that although FRIIs are less frequently found in clusters overall, those that do reside in clusters occupy similar radial locations to FRIs.

    Finally, we compared the BCGs among the two classes. We found no significant differences in the richness or $M_{500}$ of the clusters hosting FRI-BCGs and FRII-BCGs, particularly within the paired sample where host properties are matched. This suggests that once a radio galaxy resides at the centre of a cluster, its FR morphology is not strongly linked to cluster-scale environments \citep{Wan1996ApJ...467..145W}.

    In summary, although FRIs are more frequently found in galaxy clusters than FRIIs, our results show that on megaparsec scales both classes exhibit similar environmental characteristics when matched in local conditions. These findings do not rule out the jet disruption scenario, but suggest that the cluster-scale environment alone does not determine the observed morphological dichotomy. This is in agreement with \cite{deJong2024A&A...683A..23D}, who found for FRIs and FRIIs a similar trend for the space density evolution as a function of redshift for different radio luminosity bins, even though overall cluster-scale environmental conditions evolve with redshift. However, they caution that their results are limited by residual selection biases, which become more pronounced at higher redshifts.

    The results of this work are based on the adopted W24-based matching framework. Systematic effects, including incompleteness and contamination in photometric cluster catalogues, may influence the absolute values of the cluster match fractions. Our main result, however, is the relative environmental difference between FRIs and FRIIs under a common cluster selection and matching procedure.
    We tested the stability of the observed differences by repeating the analysis in narrow redshift intervals and by using mock catalogues to estimate the level of chance associations. These tests are consistent with the observed higher cluster match fraction of FRIs, as well as the luminosity- and stellar-mass-dependent trends discussed above, within the current statistical uncertainties. Looking ahead, deeper and more homogeneous spectroscopic coverage from ongoing and future surveys such as DESI \citep{DESICollaboration2024AJ....168...58D} and WEAVE-LOFAR \citep{2016smith} will provide more accurate cluster redshifts and mass estimates, enabling more precise assessments of the large-scale environments of FRI and FRII radio galaxies. Such data will be crucial for further disentangling the interplay between radio morphology and environment.

\section{Conclusion}\label{sec5}
    In this work, we have constructed a volume-limited sample and a redshift- and luminosity-matched paired sample of FRI and FRII radio galaxies from LoTSS DR2 at $z < 0.4$, and investigated their cluster-scale environments through cross-matching with an optical galaxy cluster catalogue based on the DESI Legacy Imaging Survey. Our main findings are:
    
    \begin{enumerate}
        \item Cluster association is more common among FRIs than among FRIIs. In the volume-limited sample, \(48.6 \pm 1.8\%\) of FRIs and \(30.6^{+2.5}_{-2.3}\%\) of FRIIs are associated with clusters. In the paired sample, the cluster association fractions are \( 45.6 \pm 3.1\% \) for FRIs and \( 32.6^{+3.0}_{-2.8}\% \) for FRIIs.
        \item Only \(6.7^{+9.5}_{-3.9}\%\) of luminous (\(L_{144} > 10^{26}\,\mathrm{W\,Hz}^{-1}\)) FRIIs are located in clusters in the volume-limited sample, and \(19.0^{+5.7}_{-4.6}\%\) in the paired sample, suggesting that powerful FRIIs are preferentially found in the field or poor groups.
        \item The distributions of cluster richness and $M_{500}$ as functions of radio luminosity and stellar mass are similar for FRIs and FRIIs.
        \item The radial density distributions of FRIs and FRIIs within clusters are nearly identical, with both peaking near $0.5 \times R_{500}$.
    \end{enumerate}

    Our results lead to two main conclusions. First, FRIs are more frequently found in galaxy clusters than FRIIs, which supports the scenario that dense environments promote jet disruption and favour the development of FRI morphology. Second, when both FRIs and FRIIs reside within clusters, their environmental properties, such as cluster richness, $M_{500}$, and radial distribution, are largely similar. This implies that the cluster-scale cluster environment is unlikely to be the primary driver of the FR morphological dichotomy.

    While our results are consistent with some expectations of jet disruption models, particularly the similarity in cluster environments between FRIs and FRIIs, they do not provide strong evidence regarding the role of the cluster-scale environment in shaping radio morphology. Several limitations should be acknowledged. Our cluster associations rely on projected distances and may not fully reflect the substructure or dynamical state, and both cluster richness and $M_{500}$ are relatively coarse proxies that do not capture the detailed gas conditions relevant to jet propagation. A more complete and homogeneous environmental characterisation is needed. This will be facilitated by the forthcoming richness catalogue from Croston et al. (in preparation), which covers over 240,000 LoTSS DR2 radio sources using DESI Legacy photometry and is calibrated against X-ray, SZ, and halo abundance data. The catalogue is expected to provide improved estimates of the halo-scale environment and support more precise assessments of environmental dependence in radio-loud AGNs. Future studies incorporating direct X-ray and SZ observations will be essential to probe the thermodynamic state of the surrounding gas, particularly on galaxy-group scales. Extending this analysis to higher redshifts, and combining it with deep spectroscopic and high-resolution radio imaging, will be key to disentangling the interplay between environment and intrinsic physical drivers of radio galaxy morphology.

\section*{Data availability}
    The full Table \ref{tab:10_fri_frii_clusters} is only available in electronic form at the CDS via anonymous ftp to \url{cdsarc.u-strasbg.fr} (130.79.128.5) or via \url{http://cdsweb.u-strasbg.fr/cgi-bin/qcat?J/A+A/}.

\begin{acknowledgements}
    TP acknowledges support from the CSC (China Scholarship Council)-Leiden University joint scholarship program. 
    TP sincerely thanks Judith Croston and Xiaolan Hou for their valuable suggestions for this paper. 
    We acknowledge the referee for critical and insightful suggestions, which not only enhanced this work but also provided new perspectives for our future research. 
    JMGHJdJ acknowledges support from project CORTEX (NWA.1160.18.316) of research programme NWA-ORC, which is (partly) financed by the Dutch Research Council (NWO), and support from the OSCARS project, which has received funding from the European Commission’s Horizon Europe Research and Innovation programme under grant agreement No. 101129751.

    LOFAR data products were provided by the LOFAR Surveys Key Science project (LSKSP; \url{https://lofar-surveys.org/}) and were derived from observations with the International LOFAR Telescope (ILT). LOFAR \citep{vanHaarlem2013A&A...556A...2V} is the Low Frequency Array designed and constructed by ASTRON. It has observing, data processing, and data storage facilities in several countries, which are owned by various parties (each with their own funding sources), and which are collectively operated by the ILT foundation under a joint scientific policy. The efforts of the LSKSP have benefited from funding from the European Research Council, NOVA, NWO, CNRS-INSU, the SURF Co-operative, the UK Science and Technology Funding Council and the Jülich Supercomputing Centre.

\end{acknowledgements}

\bibliography{ref} 
\bibliographystyle{aa} 
\begin{appendix}
\onecolumn

\section{Examples FRI/FRII radio galaxies and their associated clusters}\label{app}

Table \ref{tab:10_fri_frii_clusters} lists five FRI and five FRII radio galaxies together with their associated W24 clusters. The columns give the Equatorial coordinates, redshift, and radio luminosity of the radio galaxies; the cluster name, Equatorial coordinates, and redshift of the matched W24 clusters; and the projected distance to the cluster centre. The `Sample' column indicates whether a source belongs to the volume-limited sample (with value `volume'), the paired sample (with value `paired'), or both (with value `both').

\begin{table*}[ht!]
\caption{Examples of the properties of FRI and FRII radio galaxies that are associated with W24 clusters.}
\label{tab:10_fri_frii_clusters}
\centering
\small
\begin{tabular}{lcccccccccc}
\toprule
Type & $\alpha_{\mathrm{gal}}$ & $\delta_{\mathrm{gal}}$ & $z_{\mathrm{gal}}$ & $\log_{10}L_{144}$ & Cluster name & $\alpha_{\mathrm{cl}}$ & $\delta_{\mathrm{cl}}$ & $z_{\mathrm{cl}}$ & $D_{\rm proj}$ & Sample \\
 & ($\degr$) & ($\degr$) & & ($\mathrm{W\,Hz}^{-1}$) &  &  ($\degr$) & ($\degr$) &  & (kpc) &  \\*
\midrule
FRI & 2.0461 & 26.0352 & 0.1960 & 24.97 & J000808.3+260313 & 2.0345 & 26.0536 & 0.2008 & 281.83 & volume \\
FRI & 2.7131 & 19.4908 & 0.3761 & 25.37 & J001051.1+192927 & 2.7129 & 19.4908 & 0.3761 & 0.32 & both \\
FRI & 2.9325 & 18.0642 & 0.1687 & 24.79 & J001144.7+180410 & 2.9362 & 18.0695 & 0.1687 & 0.04 & volume \\
FRI & 3.3368 & 31.9422 & 0.2737 & 25.47 & J001320.9+315638 & 3.3370 & 31.9438 & 0.2767 & 0.05 & both \\
FRI & 4.2119 & 24.4920 & 0.3582 & 25.20 & J001650.3+242927 & 4.2096 & 24.4910 & 0.3582 & 0.61 & both \\
FRII & 1.3485 & 19.1416 & 0.3058 & 25.35 & J000523.0+190831 & 1.3457 & 19.1420 & 0.3072 & 0.08 & volume \\
FRII & 2.6395 & 29.5644 & 0.3259 & 25.03 & J001038.0+293714 & 2.6585 & 29.6204 & 0.3270 & 1003.68 & both \\
FRII & 3.0107 & 27.1788 & 0.3235 & 25.21 & WH-J001203.0+271038 & 3.0124 & 27.1771 & 0.3235 & 5.49 & both \\
FRII & 7.2787 & 29.8241 & 0.3453 & 25.53 & J002906.9+294930 & 7.2789 & 29.8251 & 0.3453 & 0.12 & both \\
FRII & 8.6093 & 22.0531 & 0.3488 & 26.02 & J003426.4+220305 & 8.6099 & 22.0514 & 0.3492 & 0.14 & volume \\
\bottomrule
\end{tabular}
\end{table*}
\end{appendix}
\end{document}